\documentclass[aps,pre,twocolumn,showpacs]{revtex4}
\usepackage{graphicx,amssymb,amsmath}
\usepackage{hyperref}

\begin{document}
\widetext

\title{Universal Spectral Correlations in Orthogonal-Unitary and Symplectic-Unitary Crossover Ensembles of Random Matrices}
\thispagestyle{empty}

\author{Santosh Kumar} \email{skumar.physics@gmail.com} \author{Akhilesh Pandey}\email{ap0700@mail.jnu.ac.in}

\affiliation{School of Physical Sciences, Jawaharlal Nehru University, 
New Delhi -- 110067, India}

\begin{abstract}
Orthogonal - unitary and symplectic - unitary crossover ensembles of random matrices are relevant in many contexts, especially in the study of time reversal symmetry breaking in quantum chaotic systems. Using skew-orthogonal polynomials we show that the same generic form of $n$-level correlation functions are obtained for the Jacobi family of crossover ensembles, including the Laguerre and Gaussian cases. For large matrices we find universal forms of unfolded correlation functions when expressed in terms of a rescaled transition parameter with arbitrary initial level density.
\end{abstract}

\pacs{05.45.Mt, 24.60.Lz, 73.23.$-$b, 89.70.$-$a}
\maketitle
Random matrices have found applications in a wide class of systems exhibiting broad range of behavior \cite{RM,RMP,BG,Bnkr,Wgnr}. The Gaussian ensembles of random Hermitian matrices have turned out to be particularly useful and have been studied extensively and applied successfully to many systems. In recent years, however, non-Gaussian random matrix ensembles have drawn considerable amount of attention. The reason behind this is that, apart from the mathematical interest of extending the results known for Gaussian ensembles to the non-Gaussian ones, they turn out to be of great importance in studying physical systems where these appear naturally and hence are more appropriate in modeling the system. For example the Laguerre (or Wishart) ensembles comprise non-negative definite matrices of the type $\mathcal{A}^\dag \mathcal{A}$ (or $ \mathcal{A} \mathcal{A}^\dag$) where $\mathcal{A}$ is a rectangular Gaussian random matrix \cite{Wgnr,Dys0}. Communication theory \cite{Info,FLC}, amorphous systems \cite{FFL,SMP}, and multivariate analysis of chaotic time series \cite{chaotic} are a few examples where Laguerre ensembles of random matrices appear. Jacobi ensembles are generalizations of Gaussian and Laguerre ensembles and serve, \textit{inter alia}, as models for transmission matrices in quantum transport problems \cite{Bnkr,For}. Depending on the invariance under orthogonal, symplectic and unitary transformations the ensembles are referred to as orthogonal (OE), symplectic (SE) or unitary (UE), the first two being applicable to time reversal invariant (TRI) systems and the last to time reversal non-invariant systems \cite{RM,RMP,BG,Bnkr,Wgnr}. The corresponding Gaussian ensembles are abbreviated as GOE, GSE and GUE, the Laguerre ensembles as LOE, LSE and LUE, and the Jacobi ensembles as JOE, JSE and JUE. 
Our purpose in this work is to study non-Gaussian ensembles of random Hermitian matrices which interpolate between ensembles with orthogonal invariance and unitary invariance. The crossover is governed by a parameter $\tau$ with $\tau=0$ for the orthogonal case and $\tau=\infty$ for the unitary case. We similarly study symplectic - unitary crossover, again as a function of the parameter $\tau$. These ensembles serve as models for systems with partial TRI breaking. In applications to quantum chaotic systems $\tau$ is a measure of TRI breaking, $\tau=0$ being the TRI case. There are also applications in other systems mentioned above \cite{Bnkr,FLC,FFL,SMP}. The Gaussian ensembles with the same (OE-UE and SE-UE) crossovers have been studied earlier \cite{PM} and applied to nuclear spectra \cite{FKPT} and to spectra of quantum chaotic systems  \cite{BR}. See also \cite{PS} for the same transitions in the circular ensembles.

 We consider here the problem of (OE-UE and SE-UE ) crossovers in Laguerre and Jacobi ensembles. We find that the expression for the joint-probability density (jpd) of the eigenvalues for these transitions has a generic form applicable to all crossover ensembles belonging to the Jacobi family. We give exact compact expressions for the jpd of eigenvalues and $n$-level correlation functions for finite-dimensional matrices. This includes results for the skew-orthogonal polynomials appropriate to the transitions which are generalizations of those studied earlier \cite{PG}. We also propose a method for obtaining directly the large dimensionality limits of the unfolded two-point kernels and $n$-level correlation functions. The unfolded results are found to be the same as those in Gaussian and circular cases with the rescaled transition parameter $\lambda$ now appropriate to the Laguerre and Jacobi ensembles. We stress that the results are independent of the initial level density, as the latter affects only the unfolding function of the spectrum and the rescaling of the parameter. Finally we discuss briefly applications of these ensembles to some of the systems mentioned above.
 
 We consider $N$-dimensional Hermitian matrices with eigenvalues $x_j$ where $j=1,...,N$. The eigenvalues satisfy $\infty>x_j>-\infty$, $x_j\ge 0$, and $1 \ge x_j \ge -1$ for Gaussian, Laguerre and Jacobi ensembles respectively. We define the $\tau$-dependent jpd, $\mathcal{P}(\tau)\equiv \mathcal{P}(x_1,...,x_N;\tau)$, using Dyson's Brownian motion model \cite{Dys0,Dys1,Dys3,AP}. We have $\partial \mathcal{P}(\tau)/\partial \tau=-\mathcal{L} \mathcal{P}(\tau)$, where $\mathcal{L}$ is the Fokker-Planck operator with equilibrium jpd $\mathcal{P}_{\text{eq}}\equiv \mathcal{P}_{\text{eq}}(x_1,...,x_N)$. The similarity transformation $\xi=\mathcal{P}_{\text{eq}}^{-1/2}P$ and $\mathcal{H}=\mathcal{P}_{\text{eq}}^{-1/2}\mathcal{L}\mathcal{P}_{\text{eq}}^{1/2}$ leads to the evolution equation of $\xi$ involving Calogero-Sutherland type Hamiltonians \cite{Cal, Suth} for the ensembles mentioned above \cite{AP}. The ground state of $\mathcal{H}$ is non-degenerate with zero energy so that $\mathcal{P}_{\text{eq}}$ exists. For transitions to UE, the interaction term disappears and then $\mathcal{H}$ can be expressed as sum of $N$ single-particle Hamiltonians $\mathcal{H}_x$ representing noninteracting fermions. Thus $\mathcal{H}=\sum_{j=1}^N \mathcal{H}_{x_j}-\mathcal{E}_0$. Here $\mathcal{E}_0$ is the ground state energy of the $N$-fermion system. The $\mathcal{H}_x$ are non-negative definite operators and, for the Gaussian, Laguerre and Jacobi ensembles, are respectively \cite{Dys0,Bnkr,AP,For}
\begin{equation}
\label{HG}
\mathcal{H}_x^{G}=-\frac{1}{2}\left[\frac{\partial^2}{\partial x^2}-x^2+1\right],~~~~~~~~~~~~~~~~~~~~~~~~~~~~~~~~~~~~~
\end{equation}
\begin{equation}
\label{HL}
\mathcal{H}_x^{L}=-\frac{1}{2}\left[x\frac{\partial^2}{\partial x^2}+\frac{\partial}{\partial x}-x-\frac{(2a+1)^2}{4x}+2a+2\right],
\end{equation}
\begin{eqnarray}
\label{HJ}
\nonumber
\mathcal{H}_x^{J}=-\bigg[\!\!\!\!\!\!\!\!\!&&\!\!\!(1-x^2)\frac{\partial^2}{\partial x^2}-2x\frac{\partial}{\partial x}-\frac{\{a-b+(a+b+1)x\}^2}{(1-x^2)}\\
&+&(a+b+1)\bigg],
\end{eqnarray}
where $a>-1$, $b>-1$ in (\ref{HL}) and (\ref{HJ}).
For example, $2a+1=N'-N\ge0$ in (\ref{HL}) for transitions corresponding to the Laguerre ensembles $\mathcal{A}^\dag \mathcal{A}$, where $\mathcal{A}$ is $N'\times N$ dimensional. Here the matrix elements of $\mathcal{A}\equiv\mathcal{A}(\tau)$ are independent complex 
Gaussian variables with zero mean and same variance. For OE-UE transition the real and imaginary parts are also independent and have the variances $(1+e^{-\tau})/4$ and $(1-e^{-\tau})/4$ respectively. For SE-UE transition, same properties are valid with quaternion real and quaternion imaginary parts.

Using the Fokker-Planck operator $\mathcal{L}$, we have, for any arbitrary initial jpd $\mathcal{P}(0)$,
\begin{equation}
\label{evolv}
\mathcal{P}(\tau)=e^{-\mathcal{L}\tau}\mathcal{P}(0)=\mathcal{P}_{\text{eq}}^{1/2}e^{-\mathcal{H}\tau}\mathcal{P}_{\text{eq}}^{-1/2}\mathcal{P}(0).
\end{equation}
This expression governs the transition from any arbitrary initial jpd to the jpd corresponding to UE as the final or equilibrium density and can be efficiently utilized to solve the problems of OE-UE and SE-UE transitions. In (\ref{evolv}) $\mathcal{P}_{\text{eq}}$ corresponds to the jpd of UE,
\begin{equation}
 \mathcal{P}_{\text{eq}}\propto |\Delta_N|^2 \prod_{j=1}^N w_{\text{eq}}(x_j),
\end{equation}
where $w_{\text{eq}}$ is the square of the ground state wave function of $\mathcal{H}_x$ and $\Delta_N=\prod_{j<k}(x_j-x_k)$ is the Vandermonde determinant. For the three ensembles, $w_{\text{eq}}$ is given by
\begin{equation}
 w_{\text{eq}}(x)=\begin{cases}
            e^{-x^2} & \mbox{for GUE},\\
	    x^{2a+1} e^{-2x} & \mbox{for LUE},\\
	    (1-x)^{2a+1} (1+x)^{2b+1} & \mbox{for JUE}.
 \end{cases}
\end{equation}

In the SE-UE transition $N$ is even because the eigenvalues are doubly-degenerate for $\tau=0$ (Kramers degeneracy). We consider here $N$ as even for OE-UE transition also. The initial jpd of eigenvalues is
\begin{equation}
\label{init}
\mathcal{P}(0)\propto\Delta_N~\mbox{Pf}[\mathcal{G}^{(0)}(x_j,x_k)]\prod_{i=1}^N w(x_i).
\end{equation}
Here $j, k=1,2,...,N$, $w(x)$ is the initial weight function with boundary conditions similar to those for $w_{eq}(x)$,
\begin{equation}
\mathcal{G}^{(0)}(x,y)=\begin{cases}
           \:\:\epsilon(x-y) &\mbox{for OE},\\
           -\delta'(x-y) &\mbox {for SE},
          \end{cases} 
\end{equation}
$2\epsilon(x)$ is the sign of $x$ and $\mbox{Pf}(\mathcal{B})$ is the Pfaffian of antisymmetric matrix $\mathcal{B}$. Then Pfaffian of $\mathcal{G}^{(0)}(x,y)$ along with $\Delta_N$ leads to the factor of $|\Delta_N|$ for OE, whereas for SE it gives rise to $|\Delta_{(N/2)}|^4$ along with the delta functions for the degeneracy \cite{RM, PM, PS}. 
Using (4), (5) and the expansion of Pfaffians \cite{RM}, we get the jpd for arbitrary $\tau$,
\begin{equation}
\mathcal{P}(\tau)\propto e^{\mathcal{E}_0 \tau} \Delta_N~\mbox{Pf}[\mathcal{G}^{(\tau)}(x_j,x_k)]\prod_{i=1}^N w(x_i).
\end{equation}
Here $\mathcal{G}^{(\tau)}$ is defined in terms of one-body operator $\boldsymbol{O}_x$,
\begin{equation}
\label{Gtau}
 \mathcal{G}^{(\tau)}(x,y)=\boldsymbol{O}_x\boldsymbol{O}_y~\mathcal{G}^{(0)}(x,y),
\end{equation}
\begin{equation}
\boldsymbol{O}_x=\frac{\sqrt{w_{eq}(x)}}{w(x)}e^{-\mathcal{H}_x\tau}\frac{w(x)}{\sqrt{w_{eq}(x)}}.
\end{equation}

We now introduce the skew-orthogonal polynomials $q_j^{(\tau)}(x)$, the weighted polynomials $\phi_j^{(\tau)}(x)=w(x)q_j^{(\tau)}(x)$, and the integrated functions $\psi_j^{(\tau)}(x)$. With $j,k=0,1,2,...$, we have
\begin{equation}
\label{skew}
 \int\phi_j^{(\tau)}(x)\psi_k^{(\tau)}(x)\,dx=Z_{jk},
\end{equation}
\begin{equation}
\label{psi}
 \psi_j^{(\tau)}(x)=\int\mathcal{G}^{(\tau)}(x,y)\phi_j^{(\tau)}(y)\,dy,
\end{equation}
where $Z_{jk}=-Z_{kj}$ equals 1 for $k=j+1$ with $j$ even, $-1$ for $k=j-1$ for $j$ odd, and 0 for $|j-k|\neq 1$. [The integrand in (\ref{skew}) should be antisymmetrized in the symplectic case when $w(x)$ diverges.] Note that the polynomials are skew-orthogonal because $\mathcal{G}^{(\tau)}$ is antisymmetric. For $\tau=0$, (\ref{skew}) and (\ref{psi}) give back the OE and SE definitions \cite{RM,PG}. We can show from (\ref{skew}) and (\ref{psi}) that
\begin{equation}
\mathcal{G}^{(\tau)}=\sum_{\mu=0}^\infty\left[\psi_{2\mu}^{(\tau)}(x)\psi_{2\mu+1}^{(\tau)}(y)-\psi_{2\mu+1}^{(\tau)}(x)\psi_{2\mu}^{(\tau)}(y)\right], 
\end{equation}
and then, using (\ref{Gtau}) and (\ref{skew}), we get 
\begin{equation}
\label{psi_t}
\psi_j^{(\tau)}(x)=\boldsymbol{O}_x\psi_j^{(0)}(x),~~\phi_j^{(\tau)}(x)=(\boldsymbol{O}_x^\dag)^{-1}\phi_j^{(0)}(x).
\end{equation}
These functions can be calculated using the spectral decomposition of $\mathcal{H}_x$. 

With special choices of the initial weight function we get compact expressions for $\phi_j^{(\tau)}(x)$ and $\psi_j^{(\tau)}(x)$, similar to those for $\tau=0$ \cite{PG}. The choice $w(x)=w_a(x)\equiv x^a e^{-x}$ is appropriate for the LOE-LUE transition. We find, with $\psi_0(x)$ given by (\ref{psi_t}), 
\begin{equation}
\label{Ephi_e}
\phi_{2\mu}^{(\tau)}(x)=e^{2\mu\tau}\frac{2^{a+1/2}}{\alpha_{2\mu}}w_a(x)L_{2\mu}^{(2a+1)}(2x),~~~~~~~~
\end{equation}
\begin{eqnarray}
\nonumber
\psi_{2\mu}^{(\tau)}(x)\!&=&\!e^{-(2\mu-1)\tau}\frac{2^{a+3/2}}{\alpha_{2\mu}}\frac{w_{a+1}(x)}{2\mu}L_{2\mu-1}^{(2a+1)}(2x)~~~\\
&+&\!\!\!\frac{\alpha_{2\mu-2}}{\alpha_{2\mu}}\frac{2\mu+2a}{2\mu}\psi_{2\mu-2}^{(\tau)}(x), \:\:(\mu>0),
\end{eqnarray}
\begin{eqnarray}
\label{Ephi_o}
\nonumber
&&\!\!\!\phi_{2\mu+1}^{(\tau)}(x)=e^{2\mu\tau}\frac{2^{a+1/2}}{\alpha_{2\mu}}w_a(x)\Big[e^{\tau}(2\mu+1)~~~~~~~~~~~~~\\
&&\times L_{2\mu+1}^{(2a+1)}(2x)-e^{-\tau}(2\mu+2a+1)L_{2\mu-1}^{(2a+1)}(2x)\Big],
\end{eqnarray}
\begin{equation}
\label{Epsi_o}
\psi_{2\mu+1}^{(\tau)}(x)=e^{-2\mu\tau}\frac{2^{a+3/2}}{\alpha_{2\mu}}w_{a+1}(x)L_{2\mu}^{(2a+1)}(2x).
\end{equation}
Here $L_j^{(2a+1)}(x)$ are the associated Laguerre polynomials with weight function $w_{2a+1}(x)$ and normalization $h_j^{(2a+1)}=(\alpha_j)^2=\Gamma(j+2a+2)/\Gamma(j+1)$.
The choices $w(x)=e^{-x^2/2}$ and $(1-x)^a (1+x)^b$ give compact results respectively for GOE-GUE and JOE-JUE transitions, which, for $\tau=0$, coincide with the results in \cite{PG}. For GSE-GUE, LSE-LUE  and JSE-JUE transitions, the choices $w(x)=e^{-x^2/2}$, $x^{a+1} e^{-x}$ and $(1-x)^{a+1}(1+x)^{b+1}$ give compact answers respectively.

We now evaluate the $n$-level correlation function $R_n$, defined by
\begin{equation}
 R_n(x_1,...,x_n;\tau)=\frac{N!}{(N-n)!}\int dx_{n+1}...dx_N P(\tau),
\end{equation}
where $n=1,2,...,N$. For this, we introduce the kernels $S_N^{(\tau)}(x,y)$, $A_N^{(\tau)}(x,y)$ and $B_N^{(\tau)}(x,y)$ along with their $\tau$-evolutions. We have
\begin{eqnarray}
\label{SN}
\nonumber
S_N^{(\tau)}(x,y)\!\!\!&=&\!\!\!\sum_{\mu=0}^{(N/2)-1}\left[\phi_{2\mu}^{(\tau)}(x)\psi_{2\mu+1}^{(\tau)}(y)-\phi_{2\mu+1}^{(\tau)}(x)\psi_{2\mu}^{(\tau)}(y)\right]\\
&=&\!\!\!(\boldsymbol{O}_x^\dag)^{-1} \boldsymbol{O}_y \, S_N^{(0)}(x,y),
\end{eqnarray}
\begin{eqnarray}
\nonumber
A_N^{(\tau)}(x,y)\!\!\!&=&\!\!\!\sum_{\mu=0}^{(N/2)-1}\left[\phi_{2\mu+1}^{(\tau)}(x)\phi_{2\mu}^{(\tau)}(y)-\phi_{2\mu}^{(\tau)}(x)\phi_{2\mu+1}^{(\tau)}(y)\right]\\
&=&\!\!\!(\boldsymbol{O}_x^\dag)^{-1} (\boldsymbol{O}_y^\dag)^{-1} A_N^{(0)}(x,y),
\end{eqnarray}
\begin{eqnarray}
\nonumber
B_N^{(\tau)}(x,y)\!\!\!&=&\!\!\!\sum_{\mu=(N/2)}^{\infty}\left[\psi_{2\mu+1}^{(\tau)}(x)\psi_{2\mu}^{(\tau)}(y)-\psi_{2\mu}^{(\tau)}(x)\psi_{2\mu+1}^{(\tau)}(y)\right]\\
&=&\!\!\!\boldsymbol{O}_x\boldsymbol{O}_y B_N^{(0)}(x,y).
\end{eqnarray}
Then, using Dyson's theorems \cite{RM,Dys2}, the $n$-level correlation function can be expressed as a quaternion determinant (Qdet),
\begin{equation}
R_n(x_1,...,x_n;\tau)=\mbox{Qdet}[\sigma^{(\tau)}(x_j,x_k)]_{j,k=1,..,n},
\end{equation}
where, for both transitions, $\sigma^{(\tau)}(x,y)$ is given by
\begin{equation}
\sigma^{(\tau)}(x,y)
=\begin{bmatrix}
    S_N^{(\tau)}(x,y) & A_N^{(\tau)}(x,y)\\
    B_N^{(\tau)}(x,y) & S_N^{(\tau)}(y,x)
\end{bmatrix}.
\end{equation}

For large $N$, the level density $R_1(x;\tau)=S_N^{(\tau)}(x,x)$ undergoes a smooth transition from $\tau=0$ to $\tau=\infty$ as a function of $\tau$ (or $N\tau$ in some cases). However the unfolded correlation functions $\boldsymbol{R}_n(r_1,...,r_n)=R_n(x_1,...,x_n)/R_1(x_1)\cdots R_1(x_n)$, where $x_j=x+r_j/R_1(x)$, undergo the transition for much smaller $\tau$, given by (\ref{lmd}) below. Consider $r=(x-y)R_1(x)$ where $R_1(x)\equiv R_1(x;0)$ is the level density corresponding to the initial ensemble. For large $N$ we have $\mathcal{H}_x-\mathcal{H}_y\approx 0$ and $\mathcal{H}_x+\mathcal{H}_y\approx 2 f(x)R_1^2(x)\partial^2/\partial r^2$, where $f(x)=-1/2,-x/2,-(1-x^2)$ for Gaussian, Laguerre and Jacobi cases respectively and is the factor appearing with $\partial^2/\partial x^2$ term in $\mathcal{H}_x$. Let 
\begin{equation}
\label{lmd}
\lambda=\sqrt{-\tau f(x)}R_1(x).
\end{equation} 
For large $N$,  $S_N^{(\tau)}/R_1(x)$ in (\ref{SN}) becomes independent of $\tau$ and has the limit
\begin{equation}
\label{Sr}
 S(r)=\frac{\sin \pi r}{\pi r}=\frac{1}{\pi}\int_0^\pi\cos kr\;dk
\end{equation}
for both the transitions. On the other hand $A_N^{(\tau)}/(R_1(x))^2$ and $B_N^{(\tau)}$ have the limits 
\begin{equation}
A(r;\lambda)=e^{-2\lambda^2\partial^2/\partial r^2}A(r;0),
\end{equation}
\begin{equation}
B(r;\lambda)=e^{2\lambda^2\partial^2/\partial r^2}B(r;0).
\end{equation}
We use OE and SE results \cite{RM,PM,PG} for $\lambda=0$.
Thus, we find
\begin{equation}
A(r;\lambda)=-\frac{1}{\pi}\int_0^{\pi}k\sin kr\;e^{2\lambda^2k^2}\;dk,
\end{equation}
\begin{equation}
B(r;\lambda)=-\frac{1}{\pi}\int_\pi^{\infty}\;\frac{\sin kr}{k}\;e^{-2\lambda^2k^2}\;dk,
\end{equation}
for OE-UE transitions. Similarly we obtain
\begin{equation}
A(r;\lambda)=-\frac{1}{\pi}\int_0^{\pi}\frac{\sin kr}{k}\;e^{2\lambda^2k^2}\;dk,
\end{equation}
\begin{equation}
B(r;\lambda)=-\frac{1}{\pi}\int_\pi^{\infty}\;k \sin kr\;e^{-2\lambda^2k^2}\;dk,
\end{equation}
for SE-UE transitions. For both transitions
\begin{equation}
 \boldsymbol{R}_n(r_1,...,r_n;\lambda)=\mbox{Qdet}\left[\sigma(r_j-r_k;\lambda)\right]_{j,k=1,...,n},
\end{equation}
\begin{equation}
 \sigma(r;\lambda)=\begin{bmatrix}
                    S(r) & A(r;\lambda)\\
		    B(r;\lambda) & S(r)
                   \end{bmatrix}.
\end{equation}
These results are same as those for the Gaussian and circular ensembles \cite{PM, PS}. Note that the two-level cluster function is given by $Y_2(r)=1-R_2=S^2-AB$ where $r=r_1-r_2$. The number variance and other two-point fluctuation measures derive from $Y_2$ and have been used in the study of TRI breaking in complex nuclei \cite{FKPT} and quantum chaotic systems \cite{BR}. We believe that similar transition will be obtained in the vibrational spectra of amorphous clusters \cite{FFL,SMP}. In some of these applications Laguerre ensembles are \textit{a priori} more appropriate.

In the study of quantum transport in (mesoscopic) chaotic cavities, the transmission eigenvalues $T_j$ after the transformation $x_j=2T_j-1$ are described by the Jacobi ensembles  \cite{Bnkr,For}. For JOE-JUE transition the weight function changes from $(1+x)^b$ to $(1+x)^{2b+1}$, where $2b+1=|N_1-N_2|$ and $N=\mbox{min}(N_1,N_2)$. Here $N_1$ and $N_2$ are the number of incoming and outgoing channels. Again the $\phi$ and $\psi$ functions can be found explicitly using the above operator method and all quantities of relevance for conductance fluctuations \cite{Bnkr} can be derived as a function of the TRI breaking parameter $\tau$ for arbitrary $N_1,N_2$. For example, the variance of conductance is given by $(1+e^{-2(N_1+N_2)\tau})(N_1^2N_2^2/(N_1+N_2)^4)$ for large $N_1$ and $N_2$. Our results agree with those in \cite{Bnkr} for $\tau=0,\infty$. We also remark that it should be possible to derive a long-range two-point correlation function analogous to those given in \cite{PPK} for the transitions also. In fact such expansion has already been given for transitions in the Gaussian ensembles \cite{RMP, FKPT}. Finally, our LOE-LUE transition results are directly applicable in the calculation of the Shannon capacity of a multiple input multiple output communication channel \cite{FLC} for arbitrary number of transmitting and receiving antennas.

To conclude, we have shown that the method of skew-orthogonal polynomials is valid for a large class of OE-UE and SE-UE transitions. The polynomials can be calculated from those for the initial ensembles in terms of one-body operators. Our large-$N$ results prove the universality of spectral correlations in the crossover ensembles of the Gaussian, Laguerre and Jacobi types with arbitrary initial level density in terms of a single parameter $\lambda$. (Generalizations to arbitrary initial level density can also be done for the circular ensembles.) The finite-$N$ results are useful in communication theory \cite{Info, FLC} and mesoscopic quantum transport problems \cite{Bnkr}, and the large-$N$ results are useful in quantum chaos studies \cite{FKPT, BR}. Details of this work appears elsewhere \cite{SKP}.

S. K. acknowledges CSIR, India for financial support.

\end{document}